# Uniformly Sampled Polar and Cylindrical Grid Approach for 2D, 3D Image Reconstruction using Algebraic Algorithm

**Sudhir Kumar Chaudhary[*], Pankaj Wahi, Prabhat Munshi**

*Nuclear Engineering & Technology Programme*

*Indian Institute of Technology Kanpur, Kanpur, India*

[*]*kcsudhir@iitk.ac.in*

**ABSTRACT** Image reconstruction by Algebraic Methods (AM) outperforms the transform methods in situations where the data collection procedure is constrained by time, space, and radiation dose. AM algorithms can also be applied for the cases where these constraints are not present but their high computational and storage requirement prohibit their actual breakthrough in such cases. In the present work, we propose a novel Uniformly Sampled Polar/Cylindrical Grid (USPG/USCG) discretization scheme to reduce the computational and storage burden of algebraic methods. The symmetries of USPG/USCG are utilized to speed up the calculations of the projection coefficients. In addition, we also offer an efficient approach for USPG to Cartesian Grid (CG) transformation for the visualization. The Multiplicative Algebraic Reconstruction Technique (MART) has been used to determine the field function of the suggested grids. Experimental projections data of a frog and Cu-Lump have been exercised to validate the proposed approach. A variety of image quality measures have been evaluated to check the accuracy of the reconstruction. Results indicate that the current strategies speed up (when compared to CG-based algorithms) the reconstruction process by a factor of 2.5 and reduce the memory requirement by the factor *p*, the number of projections used in the reconstruction.

## I. INTRODUCTION

Computerized tomography (CT) is used widely as a non-invasive technique for the measurement of the physical properties (attenuation constant, emissivity, refractive index, etc.) of a test specimen [1-3]. CT gained popularity in the early days, primarily as a result of its revolutionary contribution to the field of medical diagnostics. Nowadays, apart from medical applications, CT has a wide range of applications in industries and scientific research [4-6]. The unifying goal in all these applications is to search for a solution to an inverse problem to retrieve the inner information of the object using a collection of line integrals known as projections [7-9].

There are many practical situations, where we have limited projections data due to certain engineering constraints [10-12]. In these cases, the algebraic algorithms produce superior results than the standard Filtered/Convolution Back Projection method (FBP/CBP) [13-15]. In spite of their superiority over FBP, less attention was paid to algebraic algorithms because of (a) their speed of convergence and (b) the then-existing computational speed of the processors. The current day fast computers have re-energized the development of algebraic methods specially in situations where transform methods are unable to give decent acceptable images [16-18].

The speed and accuracy of the image reconstruction methods are influenced primarily by two factors, domain discretization scheme, and solution methods. The Conventional reconstruction methods employ Cartesian Grids (CG) to discretize the image space. The algebraic solution method maps the projection data space to image space using the coefficient matrices. These matrices are compute-intensive in nature and consume enormous time when the problem dimension is large.

To alleviate these issues, researchers have pre-calculated the coefficient matrices and stored them for later use in the reconstruction process [19, 20]. This method significantly reduces the computation burden of the algebraic methods but, faced serious storage issues when implemented in 3D image reconstruction. The single-view-Multiplicative Algebraic Reconstruction Technique (Sv-MART) [21] has been implemented to reduce the storage requirement. In this method, all projection lines are treated as part of a single projection. The advantage of this strategy is that it does not require storing the coefficients of the line. The role of each line is over after correcting the image. The major drawback of this method is that it necessitates a large number of projections and sensors to produce a high-quality image. The fast ray-tracing techniques [22-24] have also been employed to overcome the problem of prolonged computation time of AM algorithms. The strategy of [24] also reduces significantly the massive data storage requirement of the coefficient matrix. It performs well for large data sets of 2D cases as well as 3D data which is small in size.

Algorithms, based on the Polar Grid (PG) discretization scheme, have been presented in the literature to speed up the calculations in SPECT (single-photon emission computed tomography) and PET (positron emission tomography) reconstruction but have gained popularity only recently [25-29]. The spatial resolution of the image based on the polar grid is the highest in the center and it declines in the radial direction (i.e. PG scheme has non-uniform spatial resolution). The works reported in the literature use interpolation techniques to map polar grid image to cartesian grid image for visualization purposes [30]. The major drawback of the conventional polar grid scheme is that it requires a large number of projection data for meaningful reconstruction thus not suitable for limited data problems.

In the present work, the Uniformly Sampled Polar/ Cylindrical Grid (USPG/USCG) discretization method has been proposed to address the shortcoming of traditional polar grid-based algebraic methods. We mostly discuss the framework of 3D reconstruction using the USCG discretization scheme (theory behind the USPG is covered during the discussion of 3D reconstruction) but present the results of reconstruction for the both geometries. The USCG is constructed in such a way that each grid has the same contribution to the image domain (i.e. USCG grid has constant resolution throughout the image space). The rotational symmetry of the suggested discretization methods allows us to perform the reconstruction on-the-fly and reduces the memory need by a factor of the number of projections used in the reconstruction. An incremental approach has been used to obtain the coefficients of the lines. The MART algorithm is employed here for the image reconstruction because it is fast and it maximizes the entropy of the image in the limiting case [31].

The remainder of the paper is structured as follows: In sections II and III, we model the image reconstruction problem using the new USPG/ USCG discretization scheme. We also suggest a direct method of mapping from USPG to CG for proper visualization of the reconstructed images. Section IV is devoted to validating the proposed methods using the mathematical and experimental projections data. Lastly, in section V, we present the overall conclusions of the proposed methods.

## II. THEORETICAL FRAMEWORK

### A. *Uniformly Sampled Polar Grid*

The resolution of the image in the conventional polar grid (PG) coordinate system is the highest at the center and decreases monotonically toward the boundary of the image. The lack of detailed information near the periphery of the image space deteriorates the image quality. To overcome this scenario, we have suggested the USPG image discretization scheme. To model it, we first examine the resolution behaviour of the polar grid. In figure 1, we have shown the polar grids with and without radial lines.

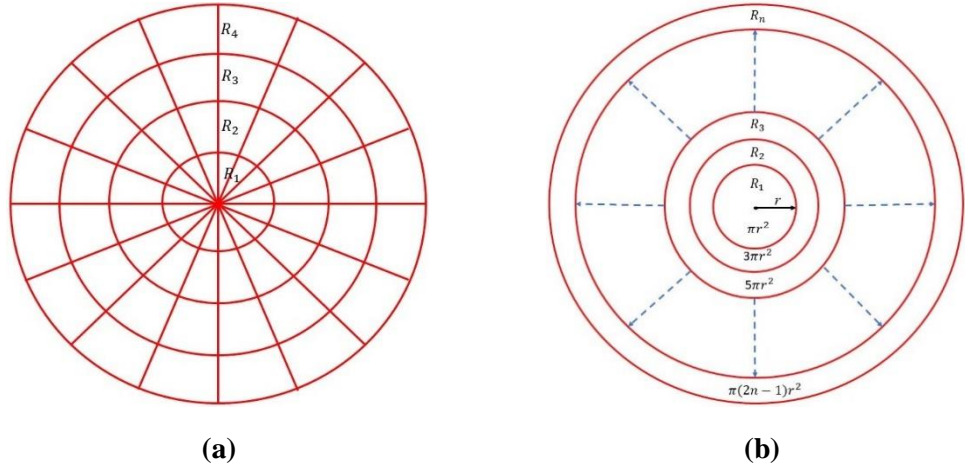

**(a)**                                **(b)**

**Figure 1.** Polar grid discretization (a) Polar grid with 4 rings and 16 radial lines (b) Polar grid with $n$ rings.

It can be seen from figure 1(a) that, in one ring, the area of pixels is the same, but it varies from one ring to another. We begin the modelling of USPG by calculating the amount of area that increases as we travel outward from the center of the coordinate system.

Each ring comprises of two concentric circles namely the inner circle and outer circle (see figure 1(b)). The radius of the inner and outer circle for the $n^{th}$ ring are $(n-1)r$ and $nr$ respectively (Rings are equally spaced with spacing $r$).

Area of the $n^{th}$ annular ring ($A_n$) then can be calculated by

$$A_n = \pi[n^2 - (n-1)^2]r^2,$$

$$A_n = \pi(2n-1)r^2. \qquad n=1,2\ldots\ldots N/2 \qquad (1)$$

where $N$ is the size of the image. Equation (1) indicates that the resolution of the ring, which is inversely proportional to $A_n$, deteriorate by the factor of odd multiples of the $\pi r^2$ ( area of the first ring) when move outward from the center. The above study suggests that we must increase the number of grids in each ring by the same amount to maintain the constant resolution. We found that ring $R_n$ have $n_g = 4(2n-1)$ number of grids to meet the criteria of constant resolution throughout the image.

Similarities between USPG and CG discretization schemes have been shown graphically in figure 2. Observation shows that circular rings of USPG and their counterpart square rings of CG have the same number of grids. In figure 2, circular rings $R_1, R_2, R_3$ and $R_4$ of USPG and square rings $C_1, C_2, C_3$ and $C_4$ of CG have 4, 12, 20 and 28 pixels respectively.

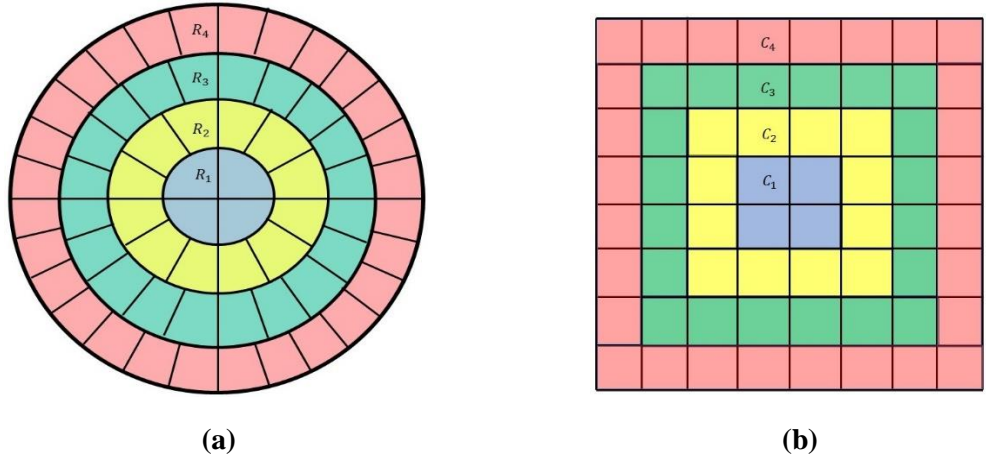

**(a)** **(b)**

**Figure 2.** Graphical representation of (a) Uniformly Sampled Polar Grid (USPG) (b) Cartesian Grid (CG).

## *B. Mapping from Uniformly Sampled Polar Grid (USPG) to Cartesian Grid (CG)*

The display panel of the system has a square-shaped pixel arrangement. Visualization of USPG reconstructed images requires the mapping from USPG to CG. Interpolation [32, 33] or iterative techniques [34] have been used previously to perform the above operation. In these methods, each pixel of the polar grid is mapped to many pixels of the cartesian grid. In the present work, we offer a simple and straightforward method for the grid transformation. This technique is based on the resemblance of USPG to CG. In the previous section, we have shown that the total number of grids in both the discretization method is the same. This feature allows us to have a perfect one-to-one correspondence (bilinear mapping) between grids of these two discretization schemes. Before proceeding to the transformation method, we provide the method of numbering and indexing for the proposed discretization grid.

Every ring in USPG has a head and a tail. The head and tail of the ring meet on the positive X-axis. We start numbering from the head and continue the numbering in anti-clockwise direction till we reach the tail (see figure 3). Ring $R_n$ have $n_g = 4(2n-1)$ grids with head element $H_n$. The head element $H_n$ of each ring can be find by using the recurrence relation,

$$H_{n+1} = H_n + 4(2n-1), \qquad (2)$$

where, *n=1,2,3………N/2* and $H_1 = 0$

| | 0 | 1 | 2 | --- | --- | --- | --- | --- | $n_g - 2$ | $n_g - 1$ |
|---|---|---|---|---|---|---|---|---|---|---|
| $R_n$ | $H_n$ | $H_n + 1$ | $H_n + 2$ | ----- | ---- | ---- | ---- | ---- | | $H_n + n_g - 1$ |
| | $-n_g$ | $-n_g + 1$ | --- | --- | --- | --- | --- | --- | -2 | -1 |

$$I = \begin{cases} i \in (0,1 \ldots n_g - 1) & + direction \\ i \in (-1, -2 \ldots -n_g + 1, -n_g) & - direction \end{cases}$$

$R_n$ is a circular list in which the tail becomes the head if the direction of indexing is reversed. We can assign index *I* to the list element either from left to right (+ direction) or right to left (- direction). A negative sign before the index shows that the direction of the indexing has been reversed. This type of indexing is very useful for fast tracing of the line through the image space. In figure 3, we have given the grid numbers of USPG in blue font and the corresponding CG numbers in black font.

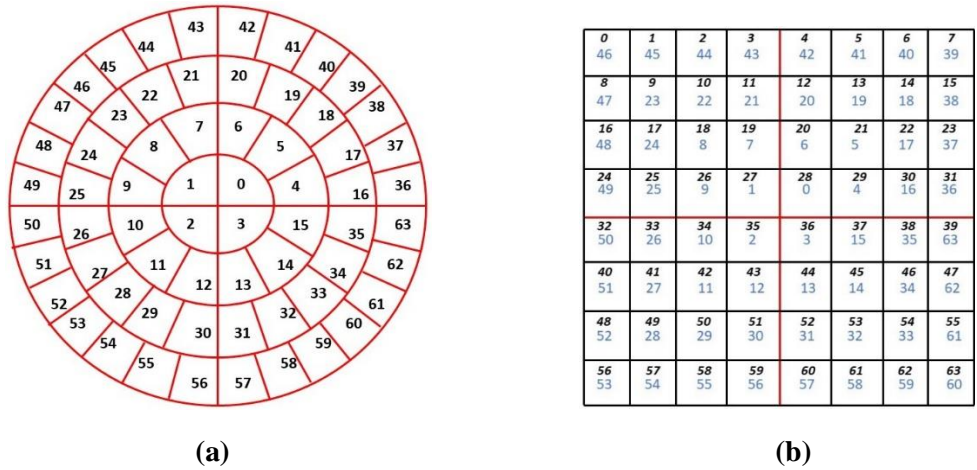

(a)                                                                                      (b)

**Figure 3.** Mapping of USPG to CG (a) Numbering strategy of uniformly sampled polar grid (b) Mapped Cartesian grid with CG numbers (black font) and corresponding USPG (blue font) numbers.

### III. MODELLING OF 3D IMAGE RECONSTRUCTION USING UNIFORMLY SAMPLED CYLINDRICAL GRID

*A. Calculation of the Intersection Points of a Line with Cylinder*

A uniformly sampled cylindrical grid (USCG) is formed by the intersection of cylindrical, radial, and axial planes. Intersections of the line with these planes are required to trace it in the image space. The procedures for calculating intersection points of the line with flat planes are simple and can be found elsewhere in the literature [24]. Here, we provide the formula for the intersection points of the line with the cylindrical plane.

The interaction of a line with a cylinder of radius $R$ has been shown in figure 4. Let $SD$ is the line between source $S$ and detector $D$. Line $SD'$ is the projection of the line $SD$ on the $XY$ plane.

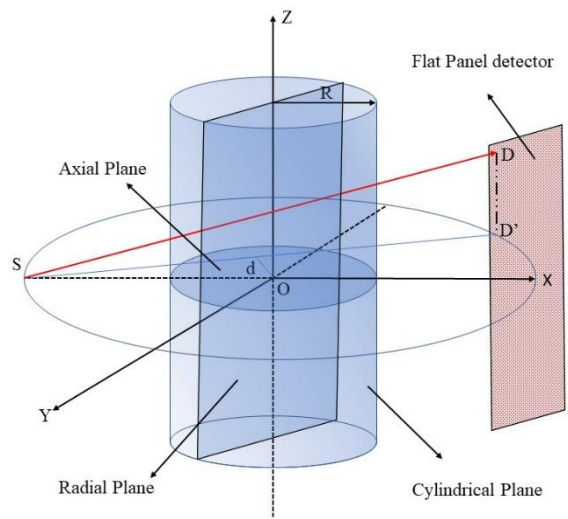

**Figure 4.** Schematic of data collection geometry for the cone beam geometry.

We first compute the distance $d$ of the line $SD'$ from the centre of the cylinder.

$$d = \sqrt{(<S,S>^2 - t^2 <S-D', S-D'>^2)}, \quad (3)$$

where,

$$t = \frac{S.(S-D')}{<S-D', S-D'>^2},$$

Line $SD$ only intersect or touch the cylinder if $d<=R$. The intersection points of line $SD$ with cylinder is given by

$$S + (t \pm k)(D-S), \quad (4)$$

where,

$$k = \sqrt{\frac{R^2 - d^2}{<S-D', S-D'>^2}}.$$

### B. Calculation of Coefficients

We have used the binary approach to calculate the coefficients of the lines. In binary approach, the grid has coefficient 1 (active grid) if it is traversed by the line otherwise coefficient is zero (dead grid). The advantage of the binary approach is that the tracing of the line does not need to perform the calculations for getting the intersection point of the line with the radial planes. Thus, the binary approach speeds up the calculation of the coefficient by O($N$).

We first calculate the intersection points of the line $SD$ with cylindrical and axial planes and then sort them according to their distances to the source point. Two consecutive intersection points $P_k$ and $P_{k+1}$ reside in the same ring and slice throughout the whole data collection process. Intersection points are the boundary points of the grid and can be shared by more than two grids. It is difficult to allocate the line coefficient to the grid on the basis of intersection points. We need a point which completely covered by a single grid. Middle point $P_m$ of the line segment $P_k P_{k+1}$ is the point which satisfies the above criteria (see figure 5)

$$P_m = \begin{cases} (P_{(k+1)x} + P_{kx})/2 \\ (P_{(k+1)y} + P_{ky})/2 \\ (P_{(k+1)z} + P_{kz})/2 \end{cases}, \quad (5)$$

In figure 6, we have shown a single slice of the USCG. The slice number $S_n$ of the line segment $P_k P_{k+1}$ can be find by dividing the $z$ coordinate of the point $P_m$ by the slice thickness $h$.

$$S_n = \left\lfloor \frac{P_{mz}}{h} \right\rfloor, \quad (6)$$

where bracket $\lfloor .. \rfloor$ converts fraction number to the floor number.

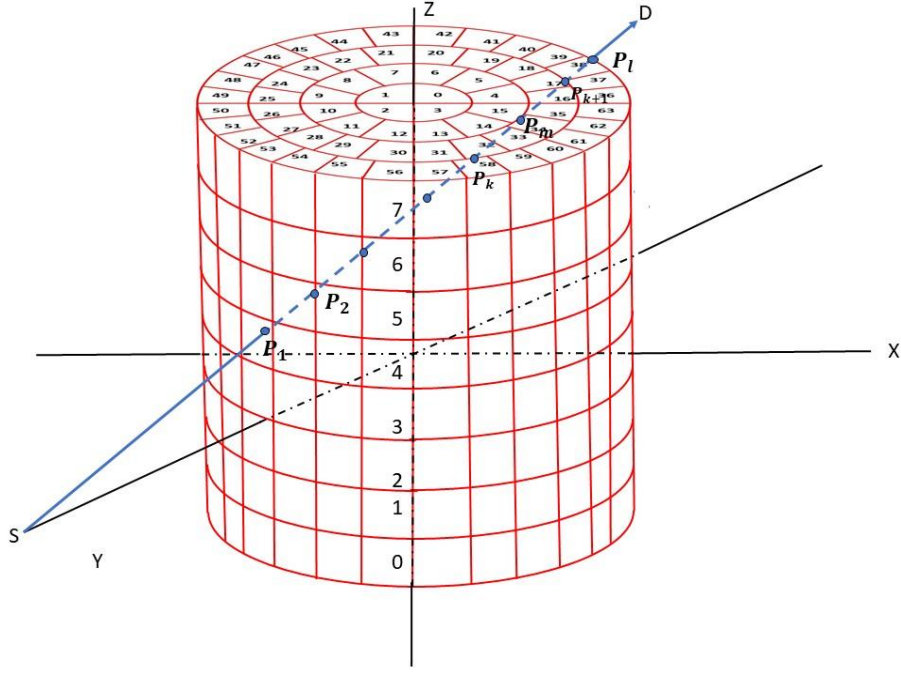

**Figure 5.** Discretization of the image space with the Uniformly Sampled Cylindrical Grid (Contains 8 slices and 4 rings).

Radius of point $P_m$ ( $d_m = \sqrt{P_{mx}^2 + P_{my}^2}$ ) and ring spacing $r_s = r$ (rings are equally spaced) decides the ring number $R_n$ of the line segment $P_k P_{k+1}$.

$$R_n = \left\lfloor \frac{d_m}{r_s} \right\rfloor, \tag{7}$$

We have the slice number $S_n$ and ring number $R_n$ of the line segment $P_k P_{k+1}$. Now we need to figure out the grids through which the line segment passes in the ring $R_n$ of the slice $S_n$.

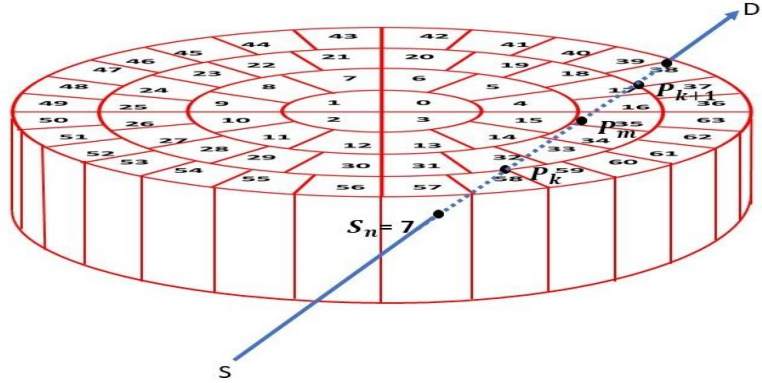

**Figure 6.** Single Slice of Uniformly Sampled Cylindrical Grid ( $S_n = 7$, $R_n$=3).

Tracing of the line segment $P_k P_{k+1}$ needs angular locations of the points $P_k$ and $P_{k+1}$. The azimuthal angles for points $P_k$ and $P_{k+1}$ are given by,

$$\varphi_{P_k} = \tan^{-1} \frac{P_{ky}}{P_{kx}} \quad and \quad \varphi_{P_{k+1}} = \tan^{-1} \frac{P_{(k+1)y}}{P_{(k+1)x}} , \tag{8}$$

Each rings have a different azimuthal step angle $\Delta\varphi_{R_n}$(angle between two consecutive radial planes of ring $R_n$)

$$\Delta\varphi_{R_n} = \frac{360}{4(2n-1)}, \tag{9}$$

Grid number $G_{P_k}$ of the point $P_k$ in the ring $R_n$ can be find as follows

$$G_{P_k} = \left\lfloor \frac{\varphi_{P_k}}{\Delta\varphi_{R_n}} \right\rfloor + H_n, \tag{10}$$

where $H_n$ is the head of ring $R_n$.

Similarly, we can find the grid location $G_{P_{k+1}}$ for the point $P_{k+1}$. Grid number $G_{P_k}$ and $G_{P_{k+1}}$ are sufficient to trace the line segment $P_k P_{k+1}$ in the ring $R_n$. The line segment $P_k P_{k+1}$ passes through all the grids which fall between grids $G_{P_k}$ and $G_{P_{k+1}}$. Now two cases arise while tracing the line segment inside the ring $R_n$ (see figure 7).

**Case 1.** The angle between the point $P_k$ and $P_{k+1}$ is less than $180^0$ (i.e. $|\varphi_{P_k} - \varphi_{P_{k+1}}| < 180^0$),

In this case, we define two quantities, line segment head $L_h$ and tail $L_t$,

$$L_h = \min(G_{P_k}, G_{P_{k+1}}), \tag{11}$$

$$L_t = \max(G_{P_k}, G_{P_{k+1}}), \tag{12}$$

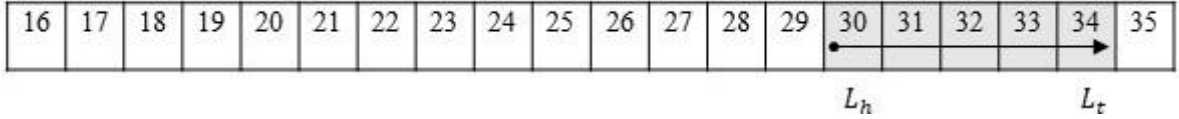

The line segment travels $(L_h - L_t + 1)$ number of grids inside the ring. Grid numbers can be given as follows,

$$G_i = L_h + i, \tag{13}$$

where, $i \in [0, L_h - L_t]$.

**Case 2.** The angle between the point $P_k$ and $P_{k+1}$ is greater than $180^0$ ($|\varphi_{P_k} - \varphi_{P_{k+1}}| > 180^0$)

In this case, point $P_k$ and $P_{k+1}$ are situated opposite to each other along the positive X axis (see Figure 7). We utilize both positive and negative index set to trace the line segment inside the ring $R_n$.

The line segment head and tail are given as follows,

$$L_h = \max(G_{P_k}, G_{P_{k+1}}), \tag{14}$$

$$L_t = \min(G_{P_k}, G_{P_{k+1}}), \tag{15}$$

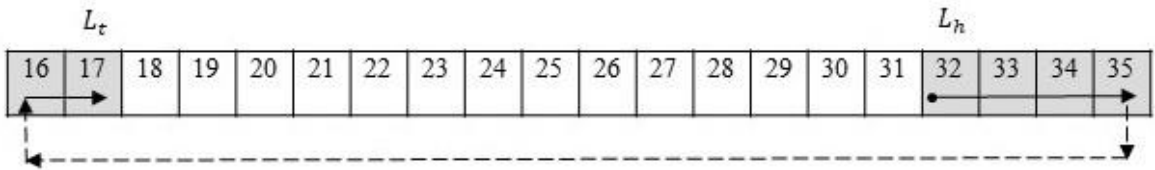

Gird numbers $G_i$ can be found,

$$G_i = R_n[L_h.index\_ + i] . \tag{16}$$

where, $L_h.index\_$ is the index of the $L_h$ in the negative direction, $i = 0,1,2...(|L_h.index\_| + L_t.index_+)$ and $L_t.index_+$ is the index of $L_t$ in the positive direction.

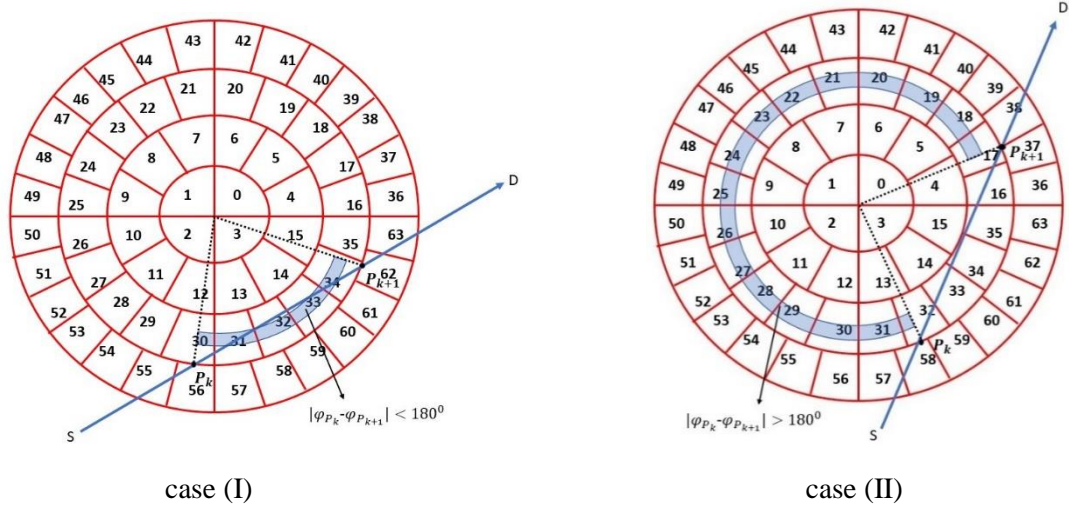

case (I)                      case (II)

**Figure 7.** Top view of slice $S_n$ to demonstrate (a) case (I) (b) case(II).

A similar process is repeated for all the line segments of the first projection. We only need to store the details of the line segments of the first projection to trace the lines of all other projections. Therefore, the proposed strategy reduces the storage burden of AM algorithms by the factor of number of projections used during the image reconstruction.

### C. Line Coefficient Calculation Algorithm based on USCG Scheme

Let the cartesian grid has dimension ($N \times N \times N$). The corresponding USCG grid then has $N$ slices, $N/2$ rings and each ring has $n_g = 4(2n - 1)$ grids where $n = 1,2..... N/2$. Every grid of USCG has radial length $r$ and slice thickness $h$. The source and detector arrangement has been rotated $p$ times for the data acquisition. The symmetries of USCG, along the transverse and axial planes, facilitate us to calculate the intersection points for only one fourth of the total lines of the first projection. The mirror images of these points provide the intersection points for the rest of the lines of the first projection. The angular position, slice number and ring number of the line segments of the first projection need to be stored for tracing the lines in all other projections. In each projection, rotation of the line around the z-axis, changes only the angular position of the line segment keeping the slice number and the ring number same as in the first projection. Angular position of any line segment of a projection can be find by simply adding the source angle to the angular position of that line segment in the first projection. Equation 10, 11, 12 and 13 or 10, 14, 15 and 16 are used to trace the line inside the image space. As the present approach requires very few steps to trace the line in the image space, we calculate the line coefficients on the fly during the reconstruction process. The following pseudo code (Algorithm-1) elucidate the process of coefficients calculation in USCG based 3D image reconstruction.

---

**Algorithm 1.** The Algorithm for Coefficients calculation based on USCG scheme

1. *Create a list $P_t$ of length $l$ (l=number of detectors)*
2. *for $i = 1 : l$*
3. *Calculate the intersection points of the line with cylindrical planes(use equation (4)) and axial planes*
4. *Sort these points according to their distances to the source S*
$$P = (P_1, P_2, ... ... ... P_k P_{k+1} ... ... ... P_n)$$

5. Store these intersection points in $i^{th}$ position of $P_t$
6. **end** for $i$
7. **for** $j = 1 : l$
8.    **for** $k = 1 :$ number of line segments of line $l$
9.       Calculate the middle point $P_m$ of the line segment $P_k P_{k+1}$ using equation (5)
10.       Use equation (6) to find the slice number of $P_k P_{k+1}$
11.       Calculate radius of point $P_m$
$$d_m = \sqrt{P_{mx}^2 + P_{my}^2}$$
12.       Use equation (7) to get ring number $R_n$ of $P_k P_{k+1}$
13.       Compute Azimuthal angles $\varphi_{P_k}$ and $\varphi_{P_{k+1}}$ for points $P_k$ and $P_{k+1}$
14.       Store each line segment information (i.e. $S_t, R_n, \varphi_{P_k}, \varphi_{P_{k+1}}$)
15.    **end** for $k$
16.    Store each line information
17. **end** for $j$
18. Step angle $\Delta\theta = 360/p$
19. **for** $i = 1: p$
20.    source position $\theta_s = p * \Delta\theta$
21.    **for** $j = 1:$ number of lines
22.       **for** $k = 1:$ number of cords of line $j$
23.          $\varphi_{P_k} = \varphi_{P_k} + \theta_s$ ; $\varphi_{P_{k+1}} = \varphi_{P_{k+1}} + \theta_s$
24.          Use equation 10 to get the grid numbers of points $P_k$ and $P_{k+1}$
25.          $G_{P_k} = \left\lfloor \frac{\varphi_{P_k}}{\Delta\varphi_{R_n}} \right\rfloor + H_n \quad G_{P_{k+1}} = \left\lfloor \frac{\varphi_{P_{k+1}}}{\Delta\varphi_{R_n}} \right\rfloor + H_n$
26.          Use equation 11 and 12 to get $L_h$ and $L_t$
27.          **if** $|\varphi_{P_k} - \varphi_{P_{k+1}}| < 180^0$
28.             **for** $l = 0 : L_h - L_t$
29.                $G[l] = L_h + l$
30.             **end** for $l$
31.          **else**
32.             Use equation 14 and 15 to get $L_h$ and $L_t$
33.             **for** $l = 0 : |L_h.index_-| + L_t.index_+$
34.                $G[l] = R_n[L_h.index_- + l\,]$
35.             **end** for $l$
36.       **end** for $k$
37.    **end** for $j$
38. **end** for $i$

### D. Reconstruction Algorithm

We have used Sparse MART(Sp-MART) algorithm [24] for finding the solution of linear equations

$$P_{p,l} = \sum_v \alpha_{p,l,v} f_v \,, \quad (17)$$

where, $P_{p,l}$ is the measured projection data of the line $l$ of projection $p$, $\alpha_{p,l,v}$ is the coefficient of line $l$ in grid $v$ and $f_v$ is the unknown field value of grid $v$. Details of Sp-MART algorithms can be find in the literature [24]. Here we briefly summarised the steps of Sp-MART algorithm.

The arbitrary field distribution $f = (f_0, f_1 \ldots \ldots f_{N^3})$ has been used to start the solution process. This distribution is used to calculate the approximate projection data $\overline{P_{p,l}}$ of line $l$ by using equation (17).

The discrepancy $\Delta P_{p,l}$ between the measured data $P_{p,l}$ and calculated data $\overline{P_{p,l}}$ assists us to get new field value $f^{new}$ for the active grids of line $l$.

$$f_v^{new} = f_v^{old}(1 - \beta \times (1 - \Delta P_{p,l})),  \quad (18)$$

where $\beta$ is the relaxation parameter which lies in the range of 0 to 2. The field values of the image space are getting updated by using the discrepancy terms $\Delta P_{p,l}$ of all the lines in all the projections. We fix the tolerance factor '$e$' and iterate the above process till the convergence criteria,

$$abs\left[\frac{f_v^{new} - f_v^{old}}{f_v^{old}}\right] \leq e . \quad (19)$$

is fulfilled.

## IV. RESULTS

### A. Reconstruction of Mathematical Phantoms

To evaluate the performance of USPG based algorithm, we have employed three different types of 2D mathematical phantoms namely soil, foam and softwood, which mimic the real life specimens. Further, we have used a standard 3D version Shepp-Logan phantom to test our USCG based framework. In both geometries, the projections data were collected at various step angles in one full rotation of the source and detector arrangements. Table 1, shows the various data collection and solution parameters involved in the reconstruction of cyber phantoms of different size, shapes and densities. Figures 8-10 show the original and the reconstructed images of soil, foam and softwood. The coronal, sagittal and axial slices of the reconstructed images of 3D version of Shepp-Logan phantom are shown in figure 11. Different image quality assessment parameters have been listed in table 2 to show the accuracy of the proposed method.

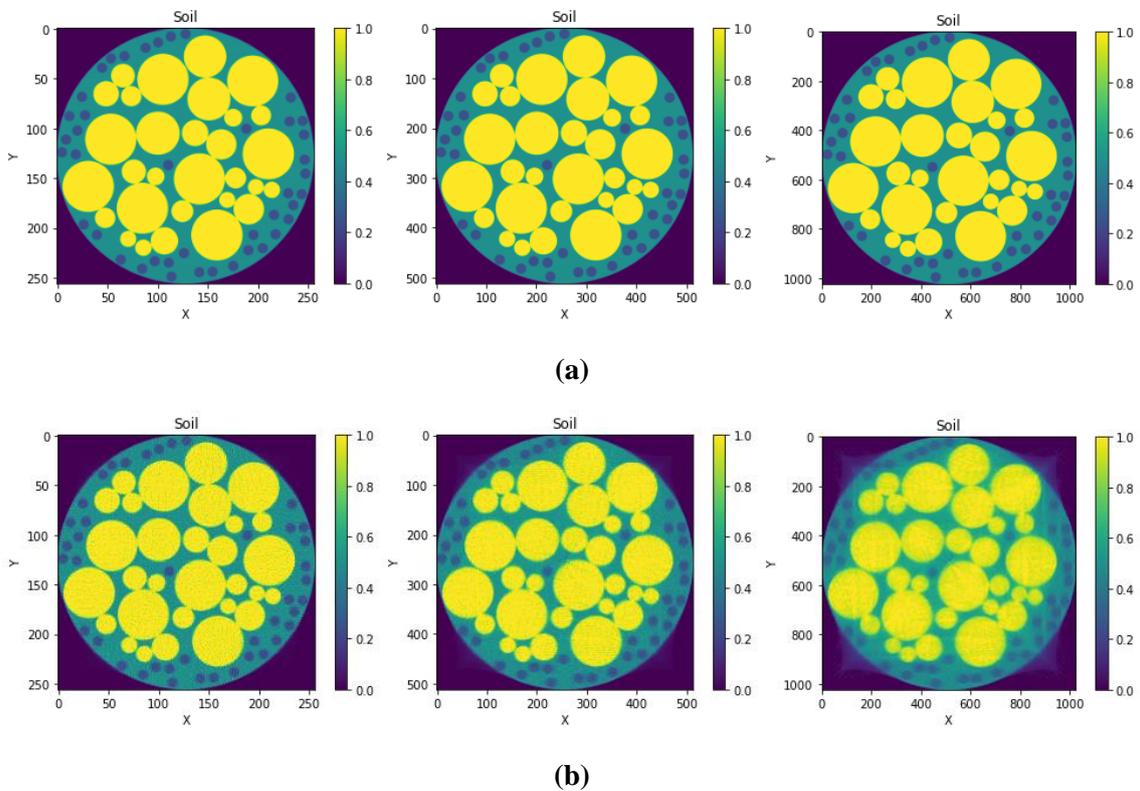

**Figure 8.** Reconstruction of 2-D mathematical phantoms of Soil of size $(256 \times 256)$, $(512 \times 512)$ and $(1024 \times 1024)$ **(a)** Original image **(b)** Reconstructed image

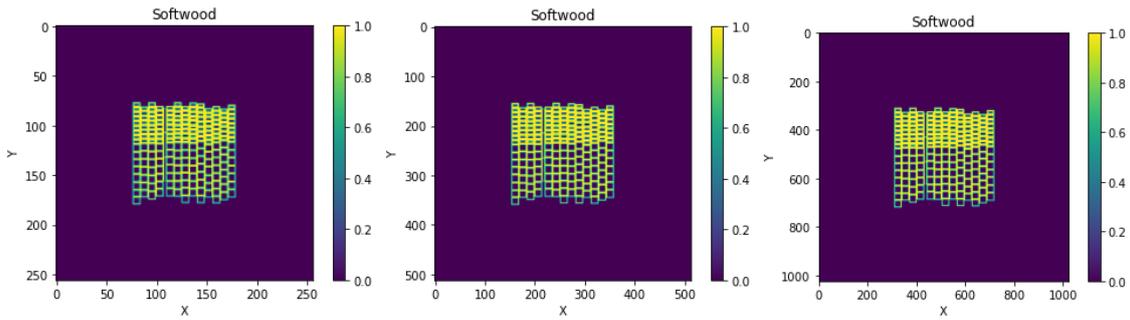
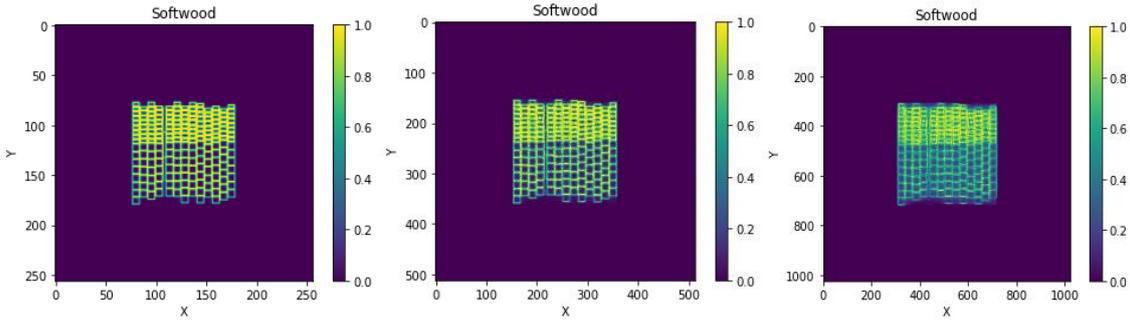

**Figure 9.** Reconstruction of 2-D mathematical phantoms of Softwood of size (256 × 256), (512 × 512) and (1024 × 1024) **(a)** Original image **(b)** Reconstructed image

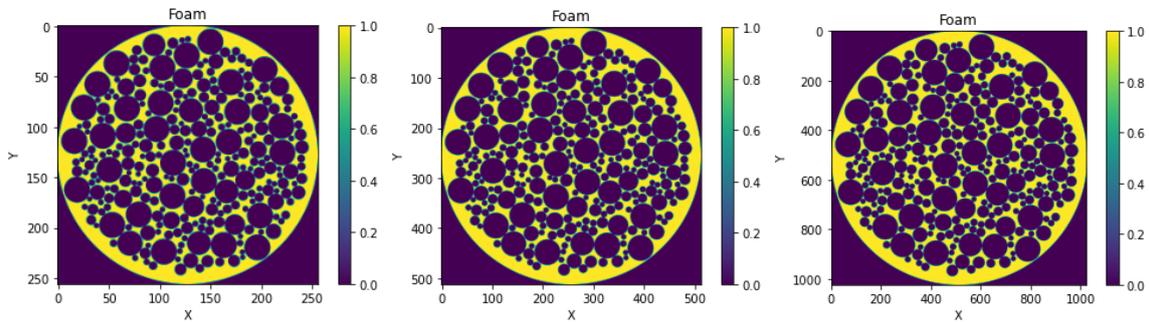
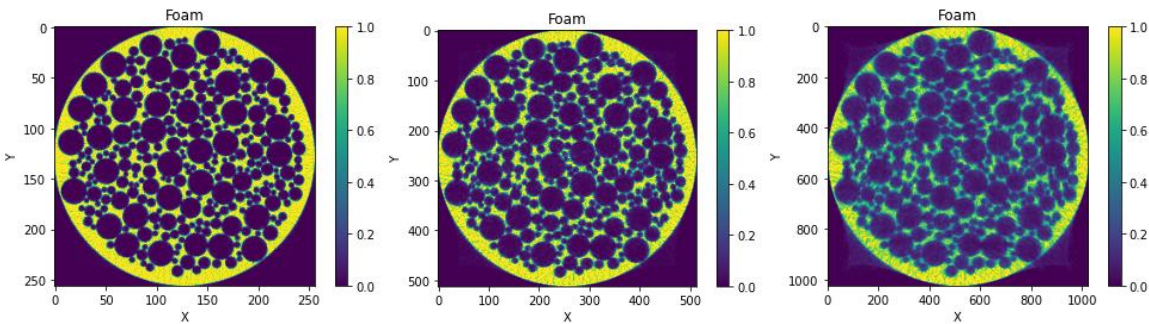

**Figure 10.** Reconstruction of 2-D mathematical phantoms of Foam of size (256 × 256), (512 × 512) and (1024 × 1024) **(a)** Original image **(b)** Reconstructed image

**Table 1.** Data collection and reconstruction parameters of mathematical phantoms

| Parameter | 2-D Phantoms | | | 3-D Shepp-Logan |
|---|---|---|---|---|
| Data collection geometry | Fan-Beam | Fan-Beam | Fan-Beam | Cone-Beam |
| Source coordinate | (-8,0) | (-8,0) | (-8,0) | (-3,0,0) |
| Center of Detector flat panel | (8,0) | (8,0) | (8,0) | (10,0,0) |
| Phantom grid size | (256 × 256) | (512 × 512) | (1024 × 1024) | (128 × 128 × 128) |
| Detector size | 101 | 101 | 101 | (101 × 101) |
| Detector spacing | 0.05 | 0.05 | 0.05 | 0.05 |
| No of projections | 50 | 50 | 50 | 70 |
| Relaxation value | 0.4 | 0.4 | 0.4 | 0.4 |

**Table 2.** Reconstruction accuracy of mathematical phantoms

| | 2-D Phantom | | | 3-D Shepp-Logan |
|---|---|---|---|---|
| | Soil | Softwood | Foam | |
| Mean Absolute Error | 0.0041 | 0.0089 | 0.0237 | 0.0013 |
| Root Mean Square Error | 0.0125 | 0.0216 | 0.0702 | 0.0035 |
| Structural Similarity Index | 0.9649 | 0.9354 | 0.8931 | 0.9898 |

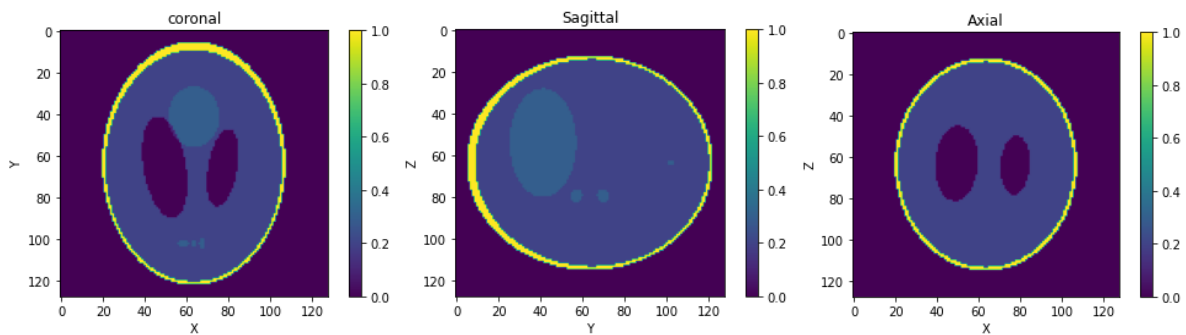

(a)

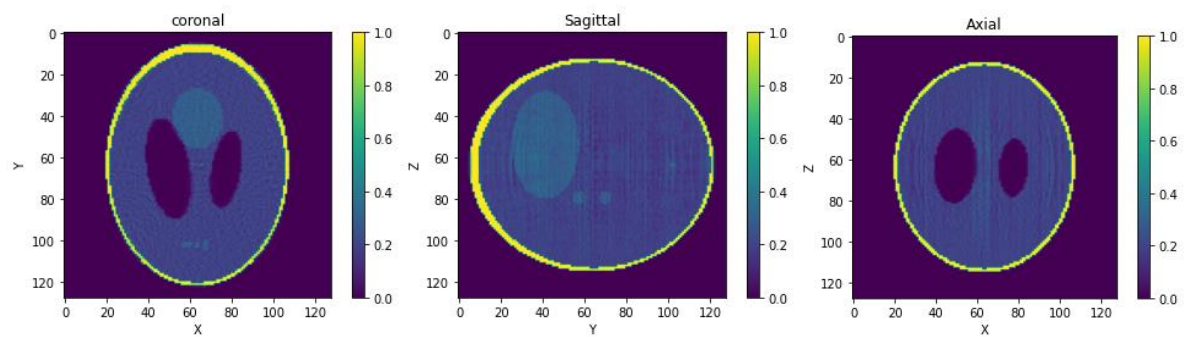

(b)

**Figure 11** Reconstruction of 3D Shepp-Logan Phantom using USCG based Sp-MART algorithm (a) Original Image (b) Reconstructed image.

All the algorithms are scripted in Python 3.8 and implemented with the system of 64 bit, 1.80 GHz, Intel® Core™ i7-8550U CPU. Total computational time of our algorithm, CG based Sp-MART and Siddon algorithm are compared in figure 12. The Siddon algorithm requires a lot of memory space for 3D reconstruction and is computationally more difficult. As a result, for the 3D reconstruction, in figure 12(b), we have displayed only the computational cost of USCG and Sp-MART algorithms. The reconstruction of 2D space of 256×256 grids needs 2.5 minutes while 3D volume of 64×64×64 grids consumes 11 minutes.

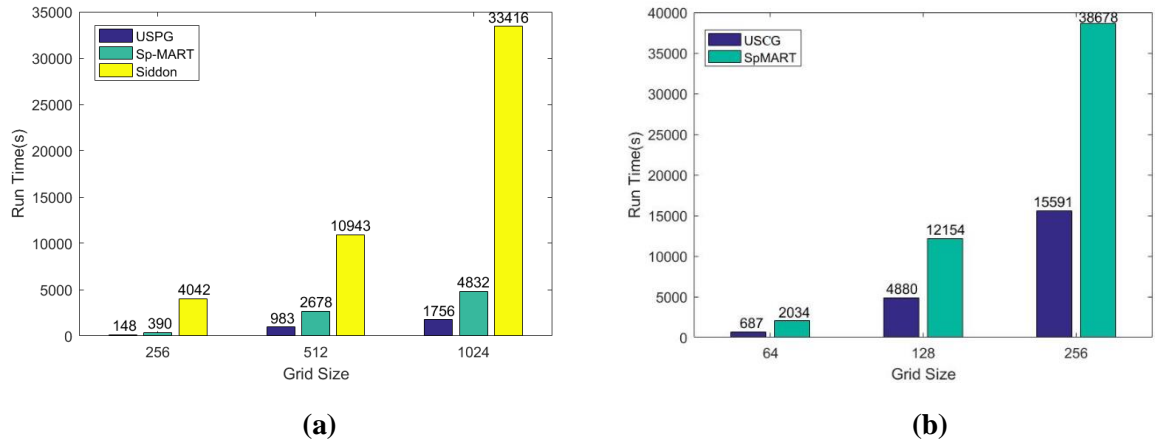

**Figure 12.** Comparison of the total run time (a) Run time for 2D image reconstruction (b) Run time for 3D image reconstruction.

## B. Reconstruction of Real Specimens

We also demonstrated the effectiveness of the proposed algorithms with the real object projection data. A Copper lump and a frog specimens are used to generate projection data using Procron X-ray CT mini machine installed at IIT Kanpur [35]. This machine has three main component- a X-ray source of 7 micron focal spot, an sample holder, and a flat panel detector arrangement (see figure 13). Detector panel has active area of $12.1 \times 12.1 \ cm^2$ and is implanted with $1024 \times 1024$ CsI detectors to count the X ray photons.

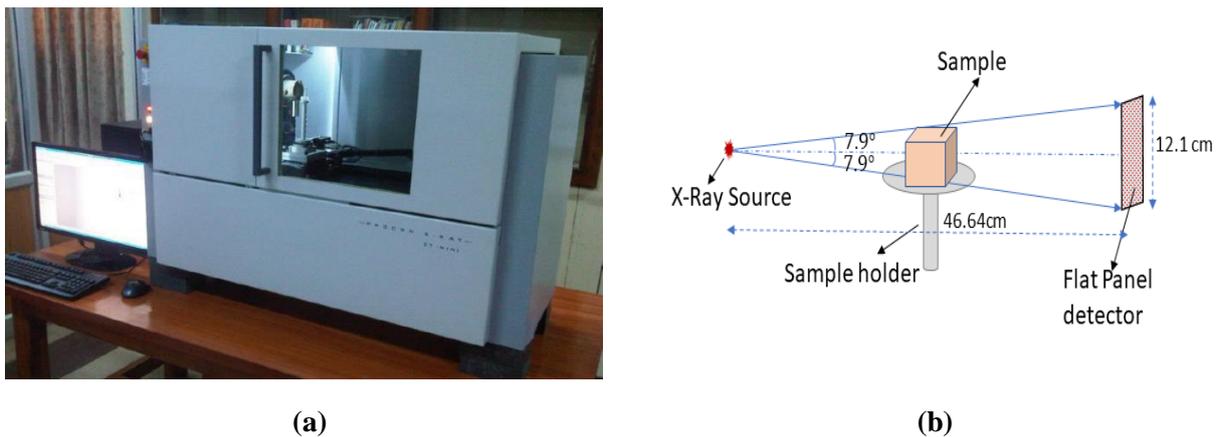

**Figure 13.** Procon X-ray CT Mini **(a)** Photograph of the scanner **(b)** schematic of the scanner

The source and object maximum admissible distance is 46.64 cm. The X-ray Source generate cone beam with cone angle of $\pm 7.9°$ on the flat panel. The object holder is able to rotate full 360° to acquire the projection data of the specimens. Various parameters involved in the reconstruction of a Copper lump and a frog are listed in table 3.

**Table 3.** Data collection and reconstruction parameters of the real specimens

| Parameter | Copper Lump | Frog |
|---|---|---|
| Source coordinate | (-8.2,0,0) | (-22.5,0,0) |
| Center coordinate of detector system | (38.4,0,0) | (24.1,0,0) |
| Reconstruction image grid size | (128 × 128 × 128) | (256 × 256 × 256) |
| Grid spacing | 0.03125 | 0.0078 |
| No of lines in each projection | 101 × 101 | 201 × 201 |
| Detector spacing | 0.121 | 0.06 |
| No of projections | 50 | 70 |
| Relaxation value | 0.4 | 0.4 |

Reconstruction slices of a Cu-lump and a frog using Feldkamp, Davis and Kress (FDK) algorithm [36] and the proposed method are shown in figure 14 and 15. The FDK algorithm requires 400 projections and more than one millions detectors to generate good quality images. We have used 120 KeV X-ray voltage and 110 micro ampere X-ray tube current for collecting experimental data of a Cu-lump. In case of frog experimental data, we have set the X-ray voltage at 110 KeV and the X-ray tube current at 105 micro ampere.

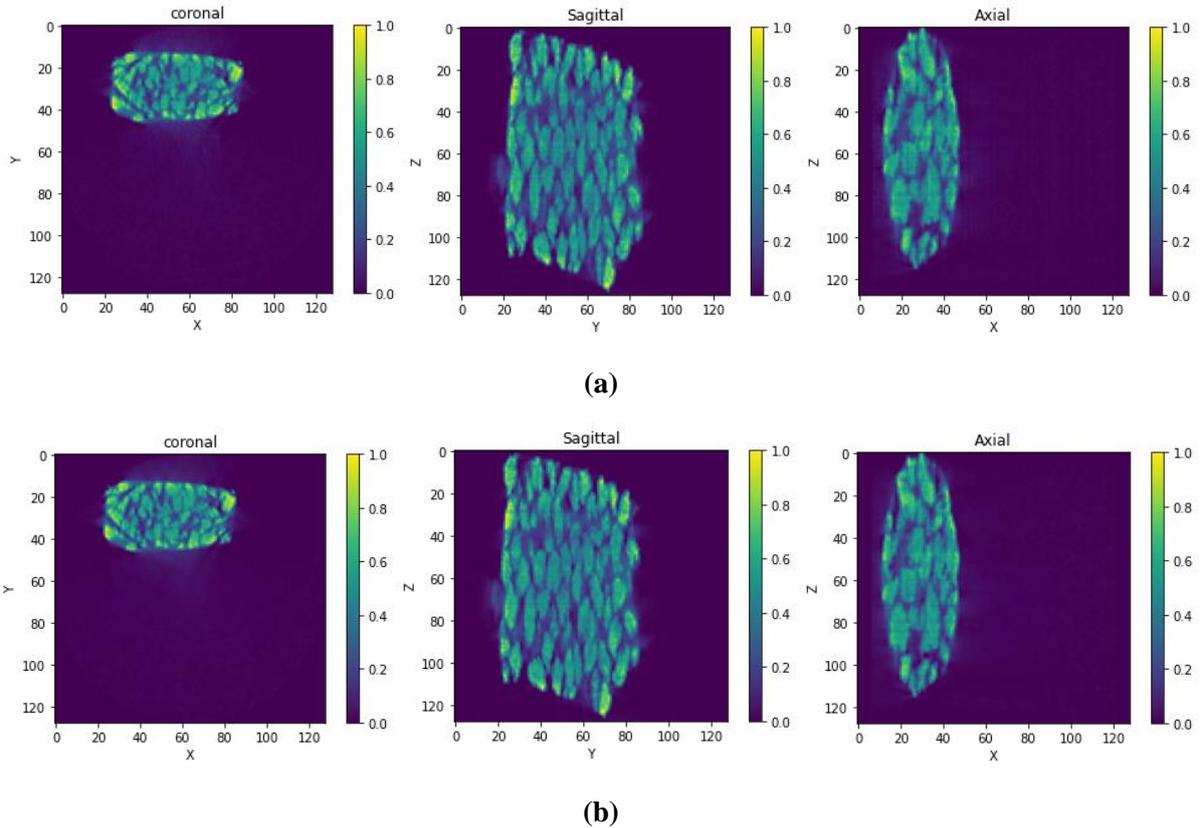

**Figure 14.** Recontructed image of Coppor Lump (central slices) (a) FDK reconstructions using 400 projection with each projection have 1.5 millions detetors (b) USCG based Sp-MART reconstruction using only 50 projections and 10 thousands detectors in each projections

Our proposed methods produces the images of the same quality with limited number of projections and detectors. Reconstruction of Copper lump incorporates 50 projections and 10 thousands detectors while frog image reconstruction employ 70 projections and 40 thousands detectors.

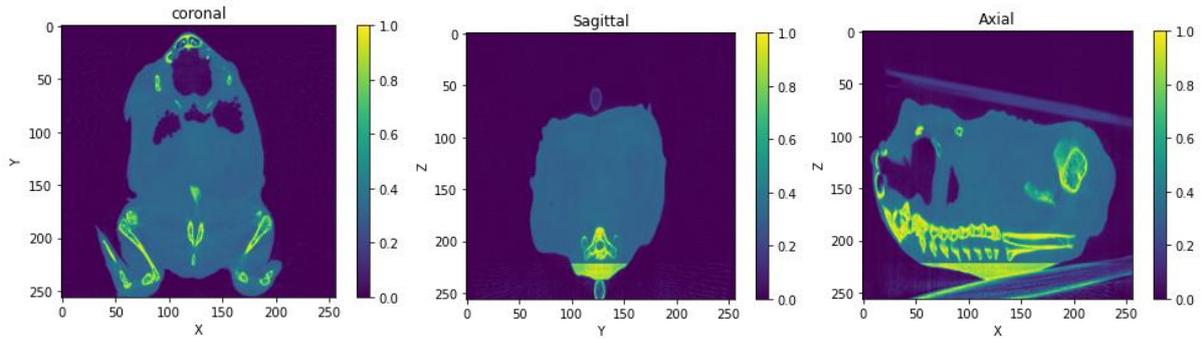

(a)

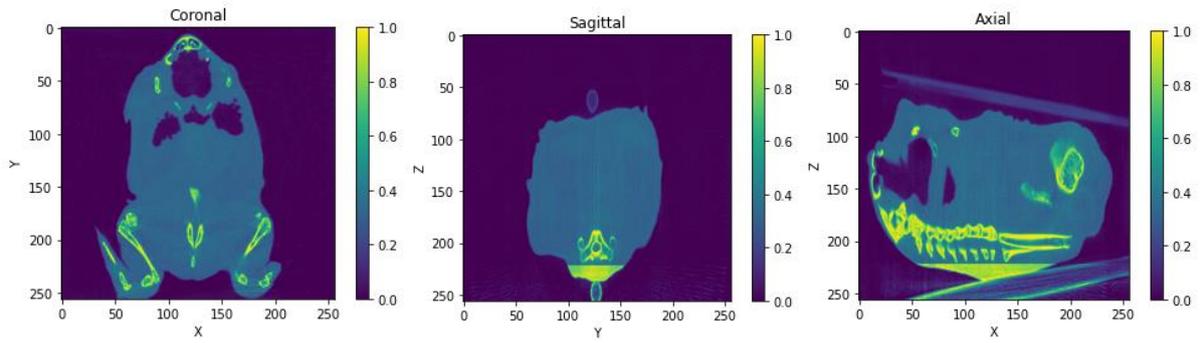

(b)

**Figure 15.** Reconstructed images of a frog (central slices) (a) FBP reconstructions using 400 projections with each projection having 1.5 millions detectors (b) USCG based Sp-MART reconstruction using only 70 projections and 40 thousands detectors in each projections.

The accuracy of the proposed method for the limited projection data was checked by considering the FDK image as the ground truth image. The intensity profiles along a line of coronal slices of the Cu-lump and frog, generated by the both algorithms, are shown in figure 16.

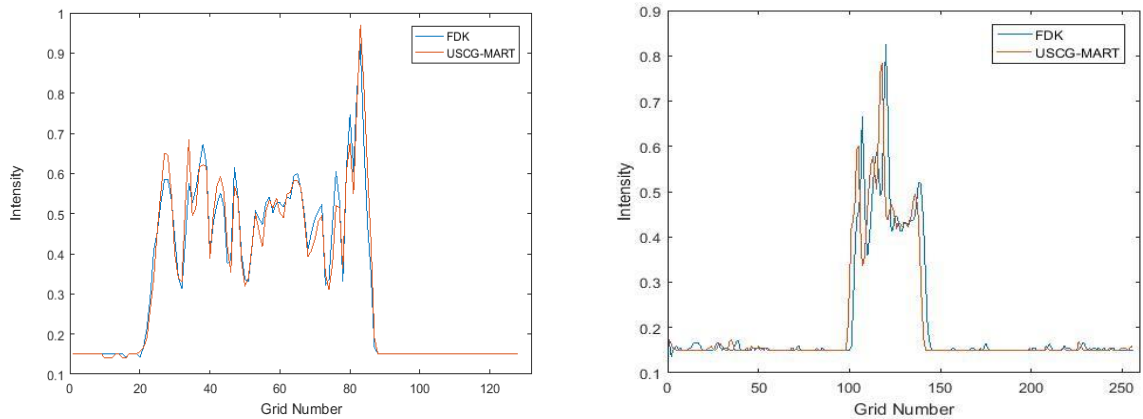

(a) (b)

**Figure 16.** Intensity profile of the coronal slice along the row 20 of the image matrix (a) Cu-lump (b) Frog

To provide more quantitative comparison, the area average, mean absolute error (MAE), root mean square error (RMSE), structural similarity index (SSIM) and sharpness measure are listed in table 4 and 5. It can be obseved from the tables that the area average of the reconstructed images by the proposed method are very close to area average of the FDK images. The iamges of Cu-Lump and frog, have small values of MAE and RMSE. Lower values of MAE and RMSE indicate strong correlation of the reconstructed image to the ground truth image. The SSIM values for both experiments are very near to 1, implying that the image pixels of FDK and proposed technique have strong inter-dependencies. In terms of sharpness also, the proposed methods perform well for the constrained image reconstruction scenario. Overall, our suggested approaches provide high-quality images with a limited number of projections and detectors, making them ideal for manufacturing the low cost CT setup.

**Table 4- Image reconstruction accuracy of a Cu-Lump (50 projections data)**

|  | Reconstruction by FDK | Reconstruction by Sp-MART |
|---|---|---|
| Area average | 0.1998 | 0.1967 |
| MAE | - | 0.0011 |
| RMSE | - | 0.0037 |
| SSIM | - | 0.9909 |
| Sharpness measure | 0.0728 | 0.0726 |

**Table 5- Image reconstruction accuracy of a frog (70 projections data)**

|  | Reconstruction by FDK | Reconstruction by Sp-MART |
|---|---|---|
| Area average | 0.2767 | 0.2736 |
| MAE | - | 0.0019 |
| RMSE | - | 0.0052 |
| SSIM | - | 0.9626 |
| Sharpness measure | 0.0897 | 0.0889 |

## V. CONCLUSIONS

In the present work, we have considered to reduce the space and time complexity of the algebraic methods using USPG/ USCG discretization scheme. Tracing the line in the traditional polar grid method requires the intersection points of the line with the cylindrical, radial, and axial planes. In the USCG strategy, we do not need to find the intersections points of the line with the radial planes. This enhances the computational speed of the line tracing process by O($N$). The only information that needs to be calculated and stored are the slice number, ring number, and angle of the line segments of the first projection view. The rotational symmetries of USCG have been used for tracing the lines of all other projections. In each projection, rotation of the line around the z-axis, changes only the angular position of the line segment keeping the slice number and the ring number same as in the first projection. Angular position of any line segment of a projection can be find by simply adding the source angle to the angular position of that line segment in the first projection. As the present approach requires very few steps to trace the line in the image space, we calculate the line coefficients on the fly during the reconstruction process. The major outcome of the present work are as follows:

(1) The issue of radial deterioration of resolution, of the conventional polar grid method, is solved by using the newly proposed USPG/ USCG discretization scheme.
(2) Only one-fourth of the total lines of the first projection need to perform calculations to obtain the intersection points with USCG. Mirror images of these points provide the intersection points for the rest of the lines.
*(3)* Most of the time-consuming steps (calculating the intersection point, sorting the arrays, rounding the numbers, finding angular position of the line segment) for computing the projection coefficients, are needed to execute only for the first projection. *As a result, the line tracing process is accelerated by O(p), where p is the number of projection views.*

(4) Projection coefficients are calculated on the fly during the reconstruction process, therefore, there is no need to store these coefficients. *It, thus, reduces the storage requirement by the factor of p.*

(5) We also presented a direct image transformation method for visualization of reconstructed images.

*(6) Overall, the speed of reconstruction is enhanced by the factor of 2.5 compared to the cartesian grid based methods.*

**ACKNOWLEDGEMENTS.** The authors would like to thank Dr. Ramesh K Aggrawal (Professor, Department of Molecular Genetics, Centre for Cellular and Molecular Biology, Hyderabad) for sharing the frog specimen and Mr. Bharat (Central Mechanical Engineering Research Institute, Durgapur) for providing the Cu-Lump sample.